\documentclass[12pt,letter]{article}

\usepackage[normalem]{ulem}
\usepackage[margin=2cm]{geometry}

\usepackage{amsfonts,amsmath,amssymb,color,graphicx,natbib, wrapfig,mwe,multirow,setspace, bbm, float, textcomp, amsthm,enumitem, subfig, algorithm,algorithmic}
\usepackage{url}

\usepackage[bottom]{footmisc}

\begin{document}
	
\title{\bfseries Accounting for total variation and robustness in profiling health care providers}

\author{Lu Xia, Kevin He, Yanming Li, John D. Kalbfleisch\footnotemark[1]  \\[4pt]
	\textit{Department of Biostatistics, University of Michigan} \\
	\textit{1415 Washington Heights, Ann Arbor, MI 48109, USA}\\[2pt]
	{$^\ast$jdkalbfl@umich.edu}
}

\markboth%
{L. XIA AND OTHERS}
{Profiling health care providers}

\date{}

\maketitle

\footnotetext[1]{To whom correspondence should be addressed.}

\begin{abstract}
	{Monitoring outcomes of health care providers, such as patient deaths, hospitalizations and hospital readmissions, helps in assessing the quality of health care. We consider a large database on patients being treated at dialysis facilities in the United States, and the problem of identifying facilities with outcomes that are better than or worse than expected. Analyses of such data have been commonly based on random  or fixed facility effects, which have shortcomings that can lead to unfair assessments. A primary issue is that they do not appropriately account for variation between providers that is outside the providers' control due, for example,  to unobserved patient characteristics that vary between providers. In this article, we  propose a smoothed empirical null approach that accounts for the total variation and adapts to different provider sizes. The linear model provides an illustration that extends easily to other nonlinear models  for survival or binary outcomes, for example. The empirical null method is generalized to allow for some variation being due to quality of care. These methods are examined with numerical simulations and applied to the monitoring of survival in the dialysis facility data.} 
	\\[0.1cm]
	
\end{abstract}


\section{Introduction}
\label{sec1}

In many instances, the quality of medical care differs considerably across providers. Large health care databases provide patient-level data that can be analyzed and summarized into provider-level measures in order to quantify and monitor such differences. Many important types of outcomes are  monitored including patient mortality, hospitalization and hospital readmission. 
Patient mortality, after adjusting for relevant patient characteristics,  is an outcome of substantial importance \citep{jarman2010hospital}. Similarly, the rate of hospital readmissions following a hospital discharge is important in monitoring hospitals and also other providers such as nursing homes and dialysis facilities since it can be indicative of the degree of coordination of post-discharge care \citep{he2013evaluating,wish2014readmission}. Provider profiling based on clinical outcomes can help patients make decisions in choosing health care providers, and also can aid overseers and payers in identifying providers whose outcomes are worse or better than a normative standard, by signaling the need for further review or to target quality improvement programs. The statistical analysis along with further review may lead to financial consequences and even suspension for providers with poor outcomes. Such evaluations require appropriate statistical methods that account for differences in patient characteristics and suitably adapt to provider size to ensure no specific groups of small or large providers are unfairly penalized in the assessment.

This paper is concerned with a large database assembled by the University of Michigan's Kidney Epidemiology and Cost Center and associated with its contacts with the Centers for Medicare \& Medicaid Services (CMS). 
A similar database is available from the United States Renal Data Systems (USRDS), in the form of standard analysis files, for use in suitable research projects. The data consist of longitudinal information on over half a million dialysis patients with end-stage renal disease (ESRD) and associated with over 6,000 dialysis facilities in the United States. These data are compiled from various sources within the CROWNWeb data system from the CMS as well as supplementary data from the Social Security master death file. The CROWNWeb data include demographic and administrative data on patients and dialysis facilities along with ongoing reporting on treatment, the Medicare claims database which can be used to determine hospitalizations, comorbidities, and other events, as well as the data  on kidney transplants in the United States. One important use of these data is to monitor dialysis facilities with respect to various quality measures including mortality, hospitalization, and readmission. In this paper, we consider the evaluation of mortality over a four-year period, accomplished through the calculation of a standardized mortality ratio (SMR) for each dialysis facility. The SMR is a risk adjusted measure which compares the observed number of deaths in a facility with the number of deaths that would be expected if the mortality rate for the center, taking account of observed patient characteristics, were the same as for a national norm \citep{reportSMR}. The measure is typically adjusted for many covariates including demographic characteristics, comorbidities at baseline, and sometimes comorbidities identified over time.  Adjusting for covariates after admission of the patient to the facility must be done with care since such comorbidities may reflect treatment choices.

Regression models for patient-level outcomes are typically used to adjust for observed patient characteristics. Methods are often based on hierarchical random effects models, in which the provider-specific effects are modeled as independent and identically distributed random variables. A typical analysis of these models is based on the empirical Bayes posterior  distribution of the provider effect where an individual provider effect is estimated by the posterior mean. We refer to this as the random-effects (RE) approach. As is well known, this analysis yields shrinkage estimates, i.e. the estimated provider effects are shrunk towards the overall average \citep{jones2011identification}. Many authors have advocated this approach; see, for example, \citet{normand1997statistical}; \citet{normand2007statistical};  \citet{jones2011identification}; \citet{ohlssen2007hierarchical}. An alternative approach, treating the provider effects as fixed, obtains the maximum likelihood estimates of the provider effects. We refer to this as the fixed-effects approach (FE). In both approaches, it is common to use the estimate and its frequency or posterior distribution to assess the sharp null hypothesis that a provider effect is identical to a national norm. 
Much of the between-provider variation is typically outside the providers' control due to inadequate risk adjustment for unobserved or observed covariates associated with the outcome that vary substantially between providers. \citet{jones2011identification} and \citet{kalbfleisch2018does} stressed that all sources of variation should be  taken into account in profiling.

In this paper, we propose empirical null methods that take into account the total variation as a robustly modeled function of provider size. The empirical null was introduced in \citet{efron2004large} and \citet{efron2010large} to account for overdispersion of Z-scores, and \citet{efron2007size} described methods to estimate a single empirical null distribution with an emphasis on controlling false discovery rates. However, these approaches that estimate and refer to a single sampling distribution do not apply immediately when providers vary markedly in size. In  \citet{kalbfleisch2013monitoring} and \citet{he2013evaluating}, this shortcoming was partially addressed  through stratification. In this article, we define an empirical null for each provider by smoothing the reference empirical null distributions. Our proposal provides a unified framework for provider profiling that encompasses various types of outcomes. We also generalize this approach to situation where a provider should be held accountable for an externally specified proportion of the total variation that is due to the quality of care.

In Section 2, we  consider issues associated with profiling in the linear model where calculations are more easily done and there is immediate comparison with familiar tests. These methods would apply directly to a quality measure that is based on normally distributed variables. In Section 3, we propose the empirical null methods with straightforward generalizations to other more complex models as well. In Section 4, we illustrate  the empirical null methods through simulation studies and analysis of the ESRD dialysis facility data with respect to monitoring standardized mortality ratios. Section 5 includes some discussion and extensions.


\section{Standard profiling methods based on the linear model}
\label{sec2}

Let $i = 1, \cdots, N$ index providers and $j = 1, \cdots, n_i$ index patients within the $i$th provider. We consider an underlying linear regression model 
\begin{equation}
Y_{ij}^* = \mu + \alpha_i + \beta^T X_{ij} + \epsilon_{ij},
\label{eq:linear_model}
\end{equation}
where $Y_{ij}^*$ represents the continuous outcome of interest with large values corresponding to poor outcomes, $\mu$ is the grand mean, $\alpha_i$ is the provider effect, $X_{ij}$ and $\beta$ are vectors of patient characteristics and regression coefficients, respectively. Our primary goal is to identify providers whose outcomes are worse than expected. 

In many instances, we assume that the  $\alpha_i$'s are random and, conditional on $X_{ij}$,
\begin{equation} \label{random_assumption}
\alpha_i \overset{iid}{\sim} N(0, \sigma^2_{\alpha}), \mathrm{\ independent\ of\ } \epsilon_{ij} \overset{iid}{\sim} N (0, \sigma_w^2).
\end{equation}
The parameters $\mu, \sigma_\alpha, \sigma_w$, and $\beta$ appear in the probability laws for all observations, and thus are estimated using all patient-level data. In what follows, we assume that the total numbers of providers, $N$, and patients, $\sum_i n_i$, are large, e.g. more than 500,000 patient records and over 6,000 dialysis facilities in our  application, so that $\mu, \sigma_\alpha, \sigma_w$ and $\beta$ can be precisely estimated. Thus, we proceed without considering the variation associated with their estimates and replace $Y_{ij}^*$'s with the risk adjusted responses, $Y_{ij}=Y_{ij}^* - \hat{\beta}^T X_{ij}$, where $\hat{\beta}$, based on fixed effects for the $\alpha_i$'s, is consistent for $\beta$. Ignoring variability in the estimation of these structural parameters is justified in large healthcare databases. In much smaller applications, one should take into account the associated variability by considering exact sampling distributions, adjustment of empirical Bayes estimates or fully Bayesian methods \citep{jones2011identification}.


\subsection{Fixed-effects, random-effects analysis, and fixed-effects with random intercept}

In a fixed-effects (FE) analysis, we consider the provider effect $\alpha_i $ in (\ref{eq:linear_model}) is a constant with a constraint, usually $\sum_{i=1}^N n_i \alpha_i = 0$, for identifiability. The maximum likelihood estimator of $\alpha_i$ is $\hat{\alpha}_i^{\mathrm{FE}} = \bar Y_i - \bar{\bar{Y}}$, where  $\bar{Y}_i = \sum_{j=1}^{n_i} Y_{ij}/n_i$ and $\bar{\bar{Y}} = \sum_{i=1}^N (n_i \bar{Y}_i) / \sum_{i=1}^N n_i$.  As noted above, $\bar{\bar{Y}}$ is an accurate estimate of $\mu$ as is $\hat{\sigma}_w$ of $\sigma_w$. The FE Z-score for testing the sharp null of $\alpha_i=0$  is 
\begin{equation}
Z_{\mathrm{FE},i} = \frac{\hat{\alpha}_i^{\mathrm{FE}} }{ \hat{\sigma}_w/\sqrt{n_i}} \approx \displaystyle \frac{\sqrt{n_i} (\bar{Y}_i - \mu)}{\sigma_w} ,
\end{equation}
which has a $N(0,1)$ distribution under the null $\alpha_i = 0$. This can be used as the reference distribution to flag unusual providers. Then,  one approach might be to flag the $i$th provider if $Z_{\mathrm{FE},i}>z_{\rho}$, where $z_{\rho}$ is the upper ${\rho}$th quantile of the standard normal distribution and $\rho$ is the nominal significance level of the related one-sided test. We use $\rho=0.05$ throughout this article.


The usual random-effects (RE) analysis of (\ref{eq:linear_model}) is based on the assumption (\ref{random_assumption}) that $\alpha_i \overset{iid}{\sim} N(0, \sigma^2_{\alpha}), \ i=1, \cdots, N$, and that the provider effects $\alpha_i$ are independent of patient characteristics $X_{ij}, \ j=1, \cdots, n_i$. The latter is an important assumption in RE analysis that is rarely met or noted. The confounding between $\alpha_i$ and $X_{ij}$ can lead to substantially biased estimates of $\beta$ and can alter the estimates of $\alpha_i$ \citep{kalbfleisch2013monitoring}. One way to address such bias is to utilize the fixed-effects estimate of $\beta$, i.e. $\hat{\beta}$, to provide an offset. As described earlier, we can obtain the risk adjusted response, $Y_{ij}$,  and then estimate $\sigma_{\alpha}$, $\sigma_w$, $\mu$ and $\alpha_i$'s assuming the model $Y_{ij} = \mu + \alpha_i + \epsilon_{ij}$.  

An empirical Bayes approach gives the approximate posterior distribution of $\alpha_i$,
\begin{equation} \label{dist_re}
\alpha_i | \{Y_{ij}\} \sim N\left(R_i (\bar{Y}_i - \mu), R_i \sigma_w^2/n_i \right),
\end{equation}
where $R_i = \sigma_{\alpha}^2 / (\sigma_{\alpha}^2 + \sigma_w^2/n_i)$. 
This approach yields an estimate $\hat{\alpha}_i^{\mathrm{RE}}= \hat{R}_i (\bar{Y}_i - \bar{\bar{Y}})$  that is shrunk toward zero by $\hat{R}_i$. These RE estimates are conditionally biased; if the $i$th provider has true effect $\alpha_i$, $E(\hat{\alpha}_i^{\mathrm{RE}} | \alpha_i) \approx R_i  \alpha_i$. The corresponding RE Z-score for a test of $\alpha_i=0$ from  (\ref{dist_re}) is
\begin{equation} \label{zscre_re}
Z_{\mathrm{RE},i} = \displaystyle \frac{\hat{R}_i  (\bar{Y}_i - \bar{\bar{Y}})}{\sqrt{\hat{R}_i  \hat{\sigma}_w^2 / n_i}} = \sqrt{\hat{R}_i } Z_{\mathrm{FE},i} \approx \displaystyle \frac{\sqrt{ R_i} (\bar{Y}_i - \mu)}{\sigma_w/\sqrt{n_i}},
\end{equation}
which has a posterior reference distribution $N(0,1)$. Thus, analogous to the flagging rule in the FE analysis,  one flags a provider  if $Z_{\mathrm{RE},i} > z_{\rho}$. Note that this  reference $N(0,1)$ is based on the posterior distribution (\ref{dist_re}); the sampling distribution for $Z_{\mathrm{RE},i}$ given $\alpha_i = 0$ is approximately $N(0, R_i)$ rather than $N(0,1)$.


The FERE approach is based on FE estimates but accounts for the total (between and within) variation. The corresponding Z-score is 
\begin{equation}
Z_{\mathrm{FERE},i} = \displaystyle \frac{\hat\alpha_i^{\mathrm{FE}}}{\sqrt{\hat{\sigma}_{\alpha}^2 + \hat{\sigma}_w^2/n_i}} \approx \displaystyle \frac{\bar{Y}_i - \mu}{\sqrt{\sigma_\alpha^2+\sigma_w^2/n_i}}. 
\end{equation}
Analogous to FE and RE methods, one flags the $i$th provider if $Z_{\mathrm{FERE},i} > z_{\rho}$. This is not a test of the sharp null hypothesis $\alpha_i = 0$, but rather assesses whether $\bar{Y}_i$ could reasonably have arisen from the model (\ref{eq:linear_model}).


\subsection{Some comments on FE, RE and FERE}

\begin{figure}[ht!] 
	\centering
	\includegraphics[width=0.6\textwidth]{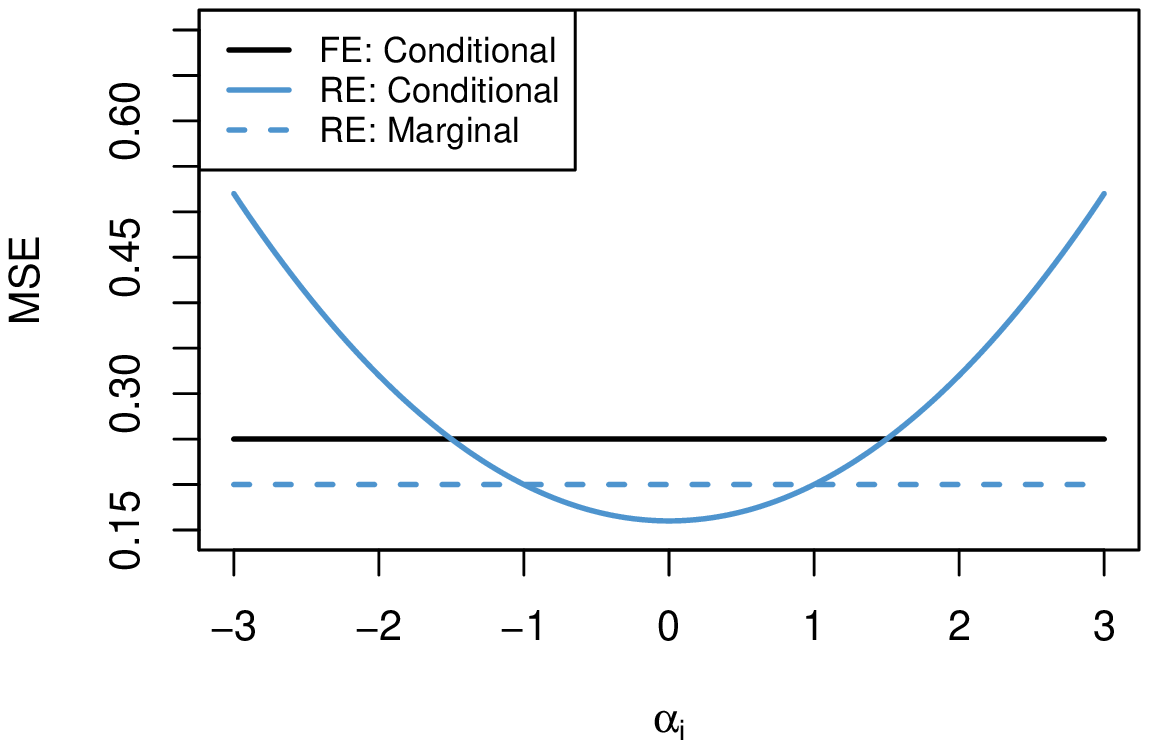}
	\caption{Conditional and marginal MSEs of the FE and RE estimates of  $\alpha_i$ in the linear model (\ref{eq:linear_model}). Conditional MSEs are calculated conditional on the true value, $\alpha_i$.  Here $\sigma_{\alpha}=1, \sigma_w = 5$ and $n_i = 100$. The marginal MSE of the FE estimate coincides with its conditional MSE.}
	\label{fig1_mse}
\end{figure}

FE and RE methods are  often used in profiling medical providers and are most appropriate when all of the variation in provider effects is due to the quality of care. Both methods make reference to a sharp null hypothesis,  $H_{0i}: \alpha_i = 0$. The RE approach is often thought to account for the variation between providers. The RE approach has been promoted largely on the basis of the well-known result that, compared to FE, it improves the overall precision of estimating the provider effects by ``borrowing information" from other providers \citep{jones2011identification,efron1973stein,louis1991assessing,krumholz2011administrative}. Thus,  the  mean squared error (MSE) of  the estimated provider effect is reduced by shrinkage, i.e. $E \left(\hat{\alpha}_i^{\mathrm{RE}} - \alpha_i \right)^2 \le E \left( \hat\alpha_i^{\mathrm{FE}} - \alpha_i \right)^2$.  This result is important in some contexts, but not in profiling.  Consider, for example, the conditional MSE, $\mathrm{MSE}^{\mathrm{met}}(\alpha_i)=E \left[\left(\hat{\alpha}_i^{\mathrm{met}} - \alpha_i \right)^2 \bigg\rvert \alpha_i  \right]$ for met=FE, RE, plotted in Fig.  \ref{fig1_mse}, where $ \sigma_{\alpha}=1, \sigma_w = 5$ and $n_i = 100$. Reduction in the overall MSE by RE is achieved by an average of substantial losses  for extreme values of $\alpha_i$ and modest gains  for more frequent values of $\alpha_i$ closer to 0. This is an example of an overall average missing important  features. For profiling, it is the extreme values that are of primary interest. For this reason among others, when the between-provider variation is entirely due to the quality of care, the FE estimates are preferred  for profiling purposes.  

Although the FE approach produces unbiased estimates of the provider effects,  the FE Z-scores are substantially overdispersed compared to the standard normal reference distribution used in testing $\alpha_i=0$  \citep{efron2004large,spiegelhalter2012statistical}. The marginal variance of the FE Z-score assuming  randomness in the $\alpha_i$'s is
\begin{equation}
Var(Z_{\mathrm{FE},i}) \approx \displaystyle \frac{\sigma_{\alpha}^2+\sigma_w^2/n_i}{\sigma_w^2/n_i} = 1 + n_i \displaystyle \frac{\sigma_{\alpha}^2}{\sigma_w^2},
\label{eq:var_zfe}
\end{equation}  
which shows the variability in FE Z-scores, especially among large providers, is much larger than the reference $N(0,1)$ distribution. The FE approach does not take into account such unexplained variation and only assesses the sharp null hypothesis of a provider effect. A large sample size can translate into more accurate estimation of the provider effect, and any deviation in $\alpha_i$ from 0  will eventually be detected by FE as $n_i \rightarrow \infty$. As a  result, the FE approach disproportionately flags  large providers even when their provider effects are  small and not clinically meaningful. Similarly, the marginal variance of $Z_{\mathrm{RE},i}$ increases linearly with $n_i$, and the same problem persists.
With a large provider size $n_i$, FE and RE are nearly identical and neither takes into account the unexplained variation between providers. 

Often, much of the variation in the $\alpha_i$'s is outside the providers' control and cannot be attributed to the quality of care. This can arise, for example, when there are unmeasured patient-level covariates, such as socio-economic status,  genetic differences and comorbidities,  that are  predictive of the outcomes but vary substantially between providers. When most of the variation between providers is outside of the providers' control, the FERE approach is preferred and its advantages have been discussed in the literature before, e.g. in the context of funnel plots which illustrate how increasing provider size affects variation in the FE estimates \citep{jones2011identification}. Providers that give rise to a large value of $Z_{\mathrm{FERE},i}$ have extreme values with reference to the population of all providers, and include all sources of variation. 

However, the usual FERE approach is not robust. When there exists a substantial proportion of providers with extreme outcomes, the between-provider variance $\sigma_{\alpha}^2$ is  overestimated, which  compromises the ability of the FERE approach to identify extreme providers. Although it is possible to develop robust estimates of $\sigma_\alpha^2$ (see, e.g.,  \citealt{koller2016robustlmm}) to improve the FERE method, generalizing them to non-linear models would be difficult.


\section{Profiling based on the empirical null}
\label{sec3}

The empirical null was introduced in \citet{efron2004large} and \citet{efron2010large} in the context of multiple testing and controlling false discovery rates. The empirical null (EN) methods in this section are designed to address overdispersion for fair assessments of all providers in a robust fashion to outliers.  These methods estimate the total variation in the FE Z-scores, $Z_{\mathrm{FE},i}, i=1,\ldots,N$,  to assess extreme values in the empirical null distributions.

\subsection{Linear model and the stratified EN approach} \label{linear_EN}

For the large majority of providers under consideration, suppose that the model (\ref{eq:linear_model}) holds,  with $\alpha_i \overset{iid}{\sim} ~ N(0, \sigma_{\alpha}^2)$. For now, we assume all providers have an equal number of patients, i.e. $n_i = n$ for $i=1, \cdots, N$.  The empirical null distribution is defined as the normal distribution with mean, $\hat{\mu}_M$, and variance, $\hat{\sigma}_M^2$, which are estimated robustly from the FE Z-scores $Z_{\mathrm{FE},i}, i=1,\ldots,N$ so as to reduce or eliminate the impact of a relativelly small number of outlying providers. This distribution is used in place of the reference $N(0,1)$ so that, for example, the $i$th provider is flagged as ``worse than expected" if $Z_{\mathrm{FE},i} > \hat{\mu}_M + z_{\rho}\hat{\sigma}_M.$

To obtain robust estimates $\hat{\mu}_M$ and $\hat{\sigma}_M^2$, we adapt the ``MLE fitting" method  \citep{efron2007size}. This is based on a mixture model in which a proportion $p_M$ of providers have Z-scores arising from the empirical null distribution, $N(\mu_M, \sigma_M^2)$, whereas the remainder, including any outliers, come from another non-null distribution.  The target quantities $\mu_M$ and $\sigma_M^2$ are estimated by maximizing the mixture likelihood so that the estimates are robust to outliers. 
More specifically, we assume that  the non-null distribution has support outside an interval $[A,B]\equiv [ \mu_M^{(0)} - \zeta_0\cdot \sigma_M^{(0)}, \mu_M^{(0)} + \zeta_0\cdot \sigma_M^{(0)}]$, where $(\mu_M^{(0)}, \sigma_M^{(0)})$ are reasonable initial estimates and $\zeta_0 > 0$ is a specified constant  (e.g. 1.2, 1.64, 1.96, or 2).  Let $I_0  = \{ i:\ Z_{\mathrm{FE},i} \in [A, B] \}$, $N_0 = |I_0|$, $N_1 = N - N_0$, $\phi_{\mu,\sigma}(z) = \displaystyle \frac{1}{\sqrt{2\pi\sigma^2}} \exp \left\{  -\displaystyle \frac{1}{2} \left( \displaystyle \frac{z - \mu}{\sigma} \right)^2  \right\}$, and $Q(\mu, \sigma) = \Phi \left( \displaystyle \frac{B - \mu}{\sigma} \right)  - \Phi \left(\displaystyle \frac{A - \mu}{\sigma} \right)$, where $\Phi(\cdot)$ is the cumulative distribution function of $N(0,1)$, and $\theta = p_M Q(\mu, \sigma) $. Then the likelihood based on the observed Z-scores is
\begin{equation} \label{en_lik}
L(\mu, \sigma, p) = \theta^{N_0} [1 - \theta]^{N_1} \prod_{i \in I_0} \displaystyle \frac{\phi_{\mu,\sigma} (Z_{\mathrm{FE},i})}{Q(\mu, \sigma)}.
\end{equation}
To obtain the MLE, $(\hat{\mu}_M, \hat{\sigma}_M, \hat{p}_M) = \displaystyle {\mathrm{argmax}}_{(\mu, \sigma, p)} L(\mu, \sigma, p)$, we proceed by computing the profile likelihood of $p_M$, which avoids solutions with $\hat{p}_M > 1$. Thus we define a grid of $K$ values for $p$ (e.g. [0.5, 1] by increment of 0.001), denoted as $\{ p^{(1)}, p^{(2)}, \cdots, p^{(K)} \}$.  For each $k = 1, 2, \cdots, K$, compute $\ell^{(k)} = \max_{\mu, \sigma} L(\mu, \sigma, p^{(k)})$ and the corresponding maximizer $(\hat{\mu}^{(k)}, \hat{\sigma}^{(k)})$. This maximization step is over a 2-dimensional parameter and is very fast using existing optimization methods (e.g. \citealt{nelder1965simplex}). Finally, find $\tilde{k} = \mathrm{argmax}_k \ell^{(k)}$, and then  $(\hat{\mu}_M, \hat{\sigma}_M, \hat{p}_M) = (\hat{\mu}^{(\tilde{k})}, \hat{\sigma}^{(\tilde{k})}, p^{(\tilde{k})})$.

An important strength of the MLE fitting approach is that the likelihood (\ref{en_lik}) is only parametric within the interval $[A, B]$, and is free from specifications of the distribution of outliers, all of which only occur outside the interval. A robust M-estimation technique can be used to obtain the preliminary estimates  \citep{huber1964robust,huber1973robust,andrews1972robust}. We use the bi-weight function, which is the default in SAS\textregistered \ ROBUSTREG procedure \citep{chen2002paper}. A stratified EN approach was previously implemented using the robust M-estimation, in \citet{kalbfleisch2013monitoring} and \citet{he2013evaluating} for time-invariant profiling, and \citet{estes2018time} for time-dynamic profiling. The MLE fitting outlined above is preferable to the robust M-estimation, since the latter can overestimate the variance with many outliers present. We have also found that moderate values of $\zeta_0$ such as $\Phi^{-1}(0.9)=1.28$ or $\Phi^{-1}(0.95)=1.64$ help avoid including outliers while allowing good estimation of $\mu_M$ and $\sigma_M$.

Next we consider a more realistic setting where the sample size $n_i$ varies across providers. Since the total variability of FE Z-scores depends on $n_i$,  we first stratify providers into a few groups based on their sample sizes, as in \citet{kalbfleisch2013monitoring} and \citet{he2013evaluating}.  For example, we may stratify the providers into three groups based on the tertiles of sample sizes $\{n_i\}$. Within  group $g$, we obtain the estimates $\hat{\mu}_{M,g}$ and $\hat{\sigma}_{M,g}$, and an empirical null distribution $N(\hat{\mu}_{M,g}, \hat{\sigma}_{M,g}^2), \ g = 1, 2, 3$. Each health care provider is compared to the empirical null corresponding to its group.  

\begin{figure}[ht]
	\centering
	\includegraphics[width=\textwidth]{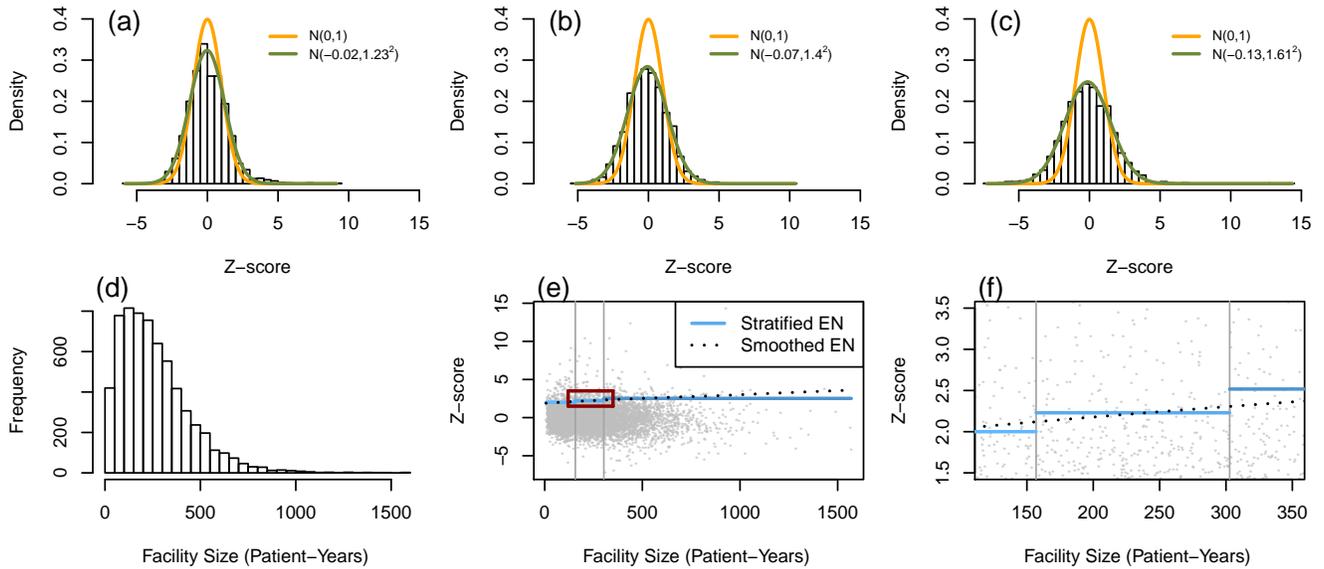}
	\caption{Histograms of FE Z-scores in dialysis mortality data (SMR), stratified by facility size tertiles into (a) small, (b) medium, and (c) large facility groups. The smooth curves represent the standard normal and the empirical null distributions. (d) Histogram of facility size in patient-years. (e) Scatter plot of FE Z-scores versus facility size along with flagging thresholds based on the stratified EN (solid blue lines) and the smoothed EN (dotted black lines). The two vertical grey lines separate facilities into three groups by the tertiles of facility size. The red square in (e) is magnified in (f). }
	\label{fig2_smr}
\end{figure}

As an illustration, consider the mortality data from the ESRD database introduced in Section 1, on patients treated at 6,363 dialysis facilities in the United States over four calendar years, from 2012 to 2015. More data descriptions can be found in Section 4.3.
Facilities are stratified into three groups, small, medium and large facilities, using facility size tertiles. The failure rates are modeled using a Cox model with covariates measuring demographic variables and comorbidities at baseline. The SMR (see Section 1) is obtained for each facility and each is converted to a one-sided Z-score based on a test of the sharp null hypothesis.  These are used in an extension of the EN methods to non-linear models as discussed in detail in Section 3.3.  
The estimated means of one-sided Z-scores are $-0.02$, $-0.07$ and $-0.13$ for the small, medium and large groups respectively, and the corresponding variance estimates are $1.23^2$, $1.40^2$ and $1.61^2$; see Fig.  \ref{fig2_smr}(a)-(c) for Z-score histograms  along with the empirical null distributions. It can be seen that the empirical null distributions are more dispersed than the standard normal, and that the variation of Z-scores increases with facility size. We use the upper 5\% quantiles in the empirical null distributions as critical values, and facilities with Z-scores larger than those in their corresponding groups are flagged as having poor outcomes. Fig.  \ref{fig2_smr}(e)-(f) visualizes the stratum-specific critical values of Z-scores in the scatter plot (solid blue segments).


\subsection{Smoothing empirical null distributions} \label{en}

The stratified empirical null approach  has some limitations, as evident from Fig.  \ref{fig2_smr}(e)-(f). First, the choice of three groups is arbitrary and a different number of strata will result in changes in the list of flagged providers. A second related problem is the discontinuity of the critical region at the stratum boundaries. Consequently, two providers near a boundary may have similar sizes and  Z-scores, yet one may be flagged and the other not, due only to the arbitrary choice of boundaries. To overcome these issues, we model the mean and variance of the empirical null distributions as smooth functions of provider size, and use robust techniques to lessen the impact of potential outliers.


To estimate the regression of the variance on the provider size, we first obtain variance estimates in each of $G$ groups defined by quantiles of provider sample size. We then regress these variance estimates on the median provider size in each group. Specifically, we proceed as follows.  Within the $g$th group, we use the local  MLE fitting described in Section \ref{linear_EN} to estimate the mean $\tilde{z}_g$ and variance $\tilde{\sigma}_g^2$ of Z-scores, and let $\tilde m_g$ represent the median size in this group, $g=1, \cdots, G$. Focusing first on the variance estimates, we fit a regression model with variance estimates $\{\tilde{\sigma}^2_g\}$ as dependent and median sizes $\{\tilde{m}_g\}$ as independent variables. Based on our empirical studies and theoretical derivations, a linear regression of the form $\gamma_0 + \gamma_1 \tilde{m}_g$ will typically suffice. An iteratively re-weighted algorithm is used to estimate $(\gamma_0, \gamma_1)$.  We set the initial estimates as the ordinary least squares estimates $(\hat{\gamma}_0^{(0)}, \hat{\gamma}_1^{(0)})$, then update with $(\hat{\gamma}_0^{(t+1)}, \hat{\gamma}_1^{(t+1)}) = \arg \min_{(\gamma_0, \gamma)} \sum_{g=1}^G \omega_g^{(t)} (\tilde{\sigma}^2_g - \gamma_0 - \gamma \tilde{m}_g)^2$ until convergence, where the weights $\omega_g^{(t)} = N_g (\hat{\gamma}_0^{(t)} +  \hat{\gamma}_1^{(t)} \tilde{m}_g)^{-2}$ and $N_g$ is the number of providers in the $g$th group. The final fitted variance values for each providers are denoted as $\hat{\sigma}_i^2,\ i = 1,\cdots, N$. We include weights in the regression to reduce the leverage of the variance estimates with large variability from large providers.

An important issue is the choice of the total number of groups $G$. A small  $G$ will not provide sufficient information for estimating regression coefficients precisely in a linear model, whereas a large $G$ results in a very small number of providers in each group and consequently unstable within group variance estimates. From our experience with the SMR example, we find it satisfactory to  choose $G$ that there are 50 to 300 providers in each group. The estimated mean and variance functions under different choices for $G$ are fairly stable, as reported in the supplementary material Section S2.


To estimate the mean Z-score as a function of provider size, we  use a weighted smoothing technique such as smoothing spline, B-spline or LOESS on data $\{ (\tilde{z}_g, \ \tilde{m}_g) \}_{g=1}^G$. The weight associated with the mean estimate $\tilde{z}_g$ in the $g$th group is inversely proportional to the variance estimate  $\hat{\sigma}_g^2$ from the iteratively re-weighted algorithm above. One can easily find existing implementations for these methods, for example, \texttt{smooth.spline()} for smoothing spline in \texttt{R}. For the provider size that falls outside the range of $\{\tilde{m}_g\}_{g=1}^G$, one may extrapolate either with the estimated linear function or with a plateau so that the mean function remains continuous and flat. The fitted Z-scores from smoothing are denoted as $\hat{Z}_i, \ i= 1, \cdots, N$. 
We propose this group-based smoothing instead of direct smoothing based on the original Z-scores, because the MLE fitting within each group is more robust against potential outliers.


Examination of the $i$th provider may proceed based on the individual empirical null distribution $N(\hat{Z}_i, \hat{\sigma}_i^2)$. Similar to other methods, the $i$th provider is flagged as worse than expected if $Z_{\mathrm{FE},i} > \hat{Z}_i + z_{\rho} \hat{\sigma}_i$, for a one-sided test with nominal level $\rho$. Smoothing provides a better approximation to the total variance of Z-scores and avoids any unfairness associated with simple stratification. When the normal linear model (\ref{eq:linear_model}) holds for all providers, this approach gives an almost identical result to the FERE approach. Fig.  \ref{fig2_smr}(e)-(f) show the critical line for flagging dialysis facilities with poor outcomes in the SMR data using the smoothed empirical null ($z_{\rho} = 1.64$ for black dotted lines).


\subsection{Extensions to non-linear models} \label{ext_nonlinear}

Many types of outcomes are monitored for quality of care, including patient hospitalizations, readmissions, transfusions and death events in the ESRD dialysis facility data.   
A hospital readmission measure within thirty days following a hospital discharge is based on a logistic model for binary outcomes, and 30-day mortality rates are analyzed in a similar manner. Hospitalizations are analyzed using a model for recurrrent events.  The empirical null approach can be readily extended to non-linear models and provides a unifying framework for profiling providers. 

We continue using the SMR example, which assesses patient death events within a health care provider.  For the $i$th provider, we denote the observed number of deaths by $O_i$. A two-stage modeling procedure is used to obtain the expected number of deaths under the assumption that patients of this provider have death events at the national average rate \citep{reportSMR}.  The following is the approach currently adopted by CMS in Dialysis Facility Compare. 

In the first stage model, the hazard function of the $j$th patient in the $i$th provider is assumed to be $\lambda_{ij}(t) = \lambda_{0i}(t)\exp\{X_{ij}^T \beta\}$, where $\lambda_{0i}$ is a provider-specific baseline hazard, and $\beta$ represents the regression coefficients associated with the observed patient-level characteristics $X_{ij}$ such as age, gender, race, BMI and a selected set of comorbidities. This stratified Cox model is fitted to the national data in order to estimate $\beta$, denoted as $\hat{\beta}$.  Stratification by providers is important to accurately estimate the within provider effects of covariates, $\beta$. 

In the second stage, the ``population-average" cumulative baseline hazard $\Lambda_0(t) = \int_{0}^t \lambda_0(u) du$ is estimated through an unstratified Cox model with an offset, $X_{ij}^T\hat\beta $, obtained from the first stage. Conditional on patient characteristics $X_{ij}$ and at-risk process $Y_{ij}(t)$, the expected number of events for the $j$th patient in the $i$th provider is calculated as
$
E_{ij} = \int_{0}^{\tau} Y_{ij}(t) \exp\{\hat{\beta}^T X_{ij}\} d\hat{\Lambda}_0(t),
$
where $\tau$ is the maximum follow-up time.  For the $i$th provider, the expected number of events is $E_i = \sum_{j=1}^{n_i} E_{ij}$ and the corresponding SMR is estimated by  $\widehat{SMR}_i = O_i / E_i$. If $\widehat{SMR}_i > 1 (<1)$, the $i$th provider experiences more (fewer) deaths than expected under the national norm given the observed characteristics of patients. Note that $E_{ij}$ is a sum of conditional expectations and is in fact the compensator in the martingale corresponding to the counting process for the individual. 

A test of the sharp null hypothesis $H_{0i}: SMR_i = 1$, where $SMR_i$ is the underlying SMR for the $i$th provider, can be obtained using a Poisson approximation whereby the number of events $O_i$ is assumed to follow a Poisson distribution with mean $E_i$. In this case, the one-sided mid p-value  is
$
p_{i} = P(X = O_i)/2 + P(X > O_i),
$
where $X \sim Poisson(E_i)$.  These can be converted to Z-scores using $Z_{\mathrm{FE},i} = \Phi^{-1} (1 - p_{i})$. By this convention, large values of $Z_{\mathrm{FE},i}$ are associated with poor outcomes.  Mid p-values are used to avoid difficulties in converting p-values to Z-scores. The  Poisson-based  p-values in this example are preferable rather than a normal approximation with Z-scores $\displaystyle \frac{O_i - E_i}{\sqrt{E_i}}$, since they are more accurate when the provider size is small. 

These FE Z-scores are consequently used to construct empirical null distributions for profiling, as introduced in Sections 3.1 and 3.2. A similar approach applies directly to other standardized measures based on other regression models, such as hierarchical logistic regression for hospital readmission.  
Instead of converting p-values, FE Z-scores could also be based on Wald statistics from fixed-effects estimates of provider effects and a test of the sharp null hypothesis. 
FE and RE methods analogous to those in the linear model are also sometimes used. RE methods tend to be complicated and subject to the same concerns as described in Section 2.2 for the linear model, and they behave similarly to FE methods in large providers and result in unreasonably higher flagging rates for large providers.

\subsection{Allowing some of the variation to be due to quality of care}

Similar to FERE, the empirical null approach presented above takes account of the total variation in the Z-scores. This is appropriate when most or all of the between-provider variation is due to incomplete risk adjustment, as opposed to the quality of care. \citet{kalbfleisch2018does} suggests introducing a value, $\lambda$, that represents the proportion of the between-provider variance that is due to incomplete risk adjustment, and holding providers accountable for a proportion of $1-\lambda$ of the between-provider variation. Note that $\lambda$ cannot be estimated on the data and must be specified based perhaps on expert opinion.

In the linear model (\ref{eq:linear_model}), we can write  $\alpha_i = \alpha_{i1} + \alpha_{i2}$, where $\alpha_{i1} \sim N(0, (1-\lambda)\sigma_{\alpha}^2)$ represents the portion of the effect due to the quality of care whereas the independent effect, $\alpha_{i2} \sim N(0, \lambda \sigma_{\alpha}^2)$, is variation that is outside the provider's control. In this case, it is natural to base profiling on an assessment of the hypothesis $H_\lambda: \alpha_{i1}=0$. Under this hypothesis, the null distribution for $Z_{\mathrm{FE}}$ is  $N \left(0, (\lambda \sigma_{\alpha}^2 + \sigma_w^2/n_i) / (\sigma_w^2/n_i ) \right)$. This is a natural generalization that connects the FE and FERE approaches discussed in Section 2, which correspond respectively to  $\lambda= 0$ and $\lambda=1$. Furthermore,  for a general $0 \le \lambda \le 1$, the null distribution for $Z_{\mathrm{FE},i}$ can also be written as  $\sqrt{1-\lambda}~N(0,1) + \sqrt{\lambda}~ N(0, 1/(1-r_i))$. Here, $r_i=\sigma_{\alpha}^2/(\sigma_{\alpha}^2+\sigma_w^2/n_i)$, the shrinkage factor defined earlier, is also referred to as the inter-unit reliability (IUR). 

Analogous to the relaxation in the linear model above, we can extend the idea of decomposing the between-provider variance to the empirical null in non-linear models.  Suppose we have obtained the empirical null distribution for a provider, $N(\hat{Z}_i, \hat{\sigma}_i^2)$. The IUR, in general, represents the proportion in the total variance that the between-provider variance takes, and can be computed in non-linear models \citep{he2018inter}. Then, for the $i$th provider with an IUR equal to $r_i$, the variance to be allowed is 
\begin{equation} 
\hat{\sigma}_{\lambda, i}^2 = \hat{\sigma}_i^2 - r_i \cdot \hat{\sigma}_i^2 (1-\lambda) = [1 - r_i(1-\lambda)] \hat{\sigma}_i^2.
\label{eq:var_lambda}
\end{equation}
Additionally, if the FE Z-scores are computed as Wald statistics or asymptotically equivalent statistics by assuming fixed provider effects, the variance in the reference distribution $\hat{\sigma}_i^2$ can be approximated by $1 / (1-r_i)$ when all between-provider variation is due to incomplete risk adjustment ($\lambda=1$), and $\hat{\sigma}_{\lambda, i}^2$ in (\ref{eq:var_lambda}) can also be written as $\hat{\sigma}_{\lambda, i}^2=1 - \lambda + \lambda \hat{\sigma}_i^2$. The new reference distribution allowing a proportion $\lambda$ of the between-provider variance is $N(\hat{Z}_i, \hat{\sigma}_{\lambda,i}^2)$. Note that the incorporation of $\lambda$ simply changes the critical value of the test based on the empirical null.

More generally, we might elicit a prior distribution $f_{\lambda}(\lambda)$ for $\lambda$ that reflects experts' uncertainty about its value. In this case, we might gauge a provider's performance by comparing its FE Z-score to the marginal empirical null distribution. Since the distribution of the data does not depend on $\lambda$, the posterior distribution is the same as the prior. The marginal empirical null distribution has density $\int_{0}^1 f_{\mathrm{EN}}(z | \lambda) f_{\lambda}(\lambda) d\lambda$, where $f_{\mathrm{EN}}(z|\lambda)$ is the density of  $N(\hat{Z}_i,\hat \sigma_{\lambda,i}^2)$. This distribution can be approximated with Monte Carlo methods that draw random samples from  $f_{\lambda}(\lambda)$ and then $f_{\mathrm{EN}}(z | \lambda)$. Given $(\hat{Z}_i, \hat{\sigma}_{i}^2)$, the mean of the marginal distribution is simply $\hat{Z}_i$ and the variance is obtained by substituting $\lambda$ with its prior mean in (\ref{eq:var_lambda}). Depending on the prior, the marginal can have heavier tails than a normal distribution. Detailed discussion is presented in the Supplementary Material Section S3.


\section{Numerical studies}

\subsection{Simulation in linear regression models}

We first restrict all providers to have the same sample size, and compare the probability that providers give rise to a signal under different approaches. We assume the true model (\ref{eq:linear_model}), with $\mu=0$, $\beta=0$, $\epsilon_{ij} \sim N(0,16)$,  $\alpha_i \sim N(0,1)$ for $i \ge 2$ and $\alpha_1$ fixed at a value varying from 0 to 3.5. We simulate $N=200$ providers of size $n_i = n$ for all $i$, where $n=10,25,50,100$. The simulation is repeated 1000 times. For one-sided tests, providers whose Z-scores exceed the corresponding critical value $\Phi^{-1}(0.95)=1.64$ are flagged as worse than expected.

\begin{figure}[th]
	\centering
	\includegraphics[width=0.8\textwidth]{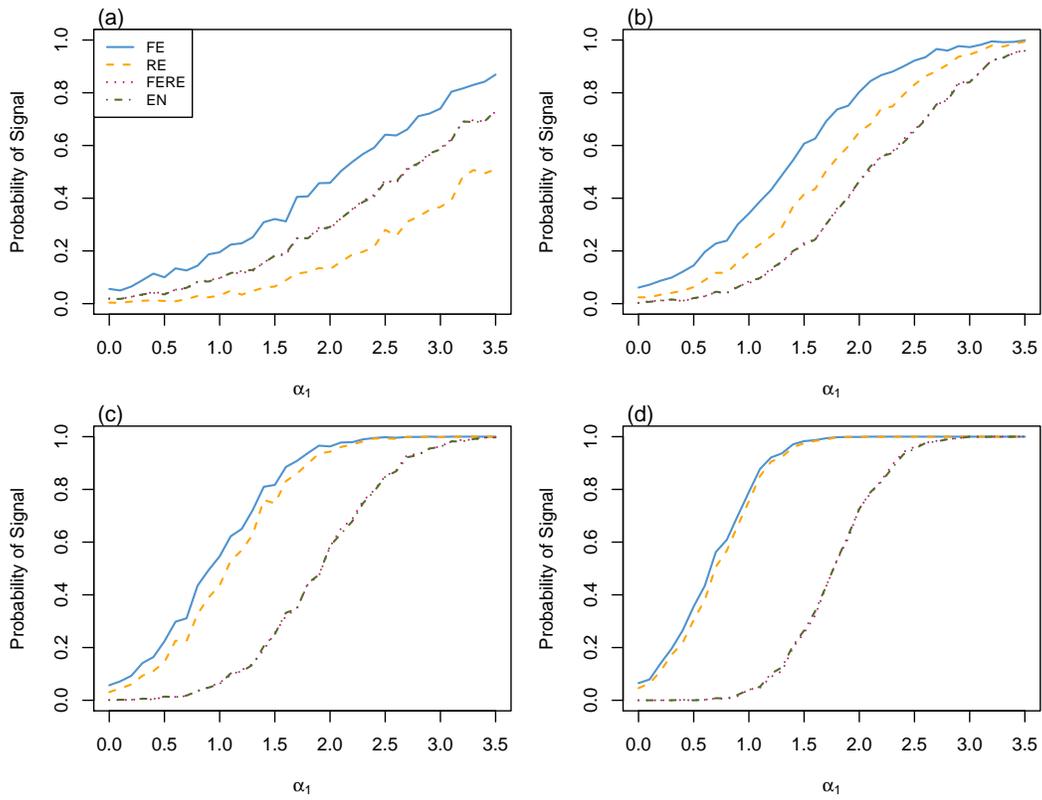}
	\caption{Estimated probability of signaling provider 1 as worse than expected under the linear model (\ref{eq:linear_model}), with all providers having sample size (a) $n=10$, (b) $n=25$, (c) $n=50$, and (d) $n=100$. The x-axis represents the fixed value of $\alpha_1$.}
	\label{fig3_lin_noout}
\end{figure}

Fig. \ref{fig3_lin_noout} shows the estimated probabilities of signaling provider 1 for FE, RE, FERE and EN approaches. The FE approach flags provider 1 with the highest probabilities in all cases. As expected, the difference between FE and RE diminishes for a large sample size $n$. Without outlying providers, FERE and EN result in almost identical probabilities of signaling, reflecting their asymptotic equivalence in this setup. For a large sample size, e.g. $n = 100$, standard FE and RE methods signal provider 1 with moderate to high probability even with relatively small values of $\alpha_1$, say $\alpha_1=0.5$ or $1$. These values of $\alpha_1$ are well within the range of variation expected for $\alpha_i$ under the true model. On the other hand,  FERE and EN allow for this variation and do not signal with  high probability until $\alpha_1>2.0$, that is until the effect is in the tail of the distribution of provider effects ($\sigma_\alpha=1$). It should be noted that the exact probability of flagging can be easily calculated and plotted for all methods except EN. We present the empirical probability in all cases to facilitate fair comparison.

\begin{figure}[th]
	\centering
	\includegraphics[width=0.8\textwidth]{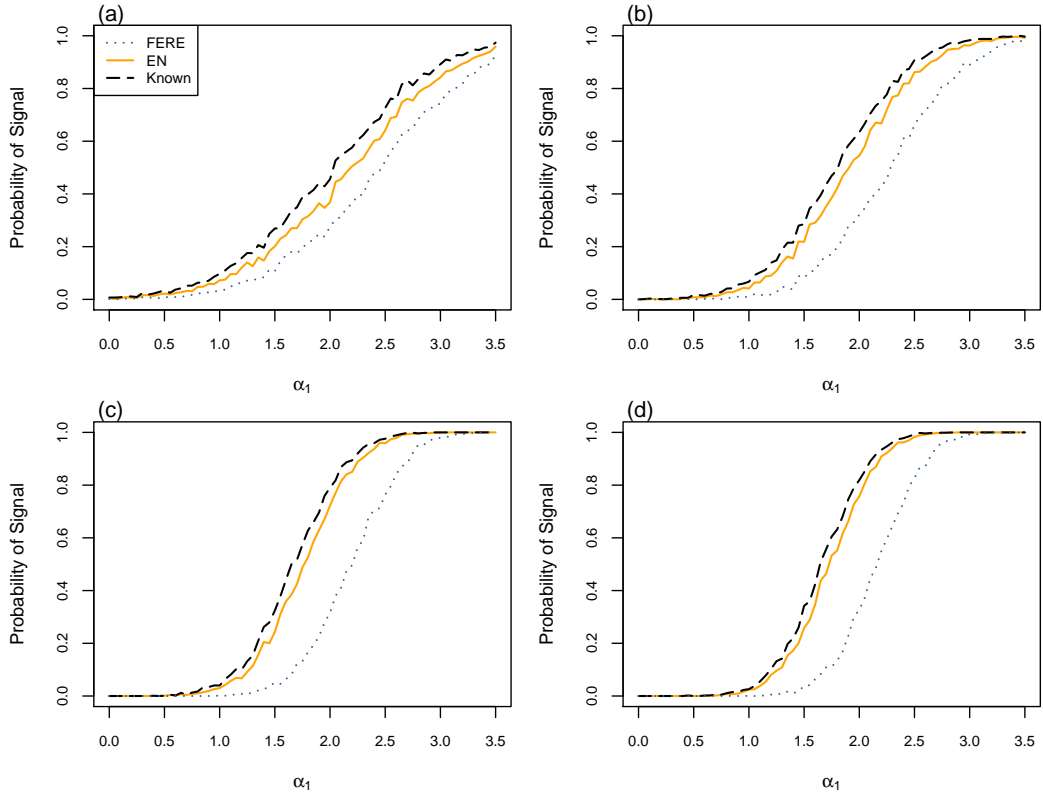}
	\caption{Estimated probability of signaling provider 1 as worse than expected under the linear model (\ref{eq:linear_model}), with 5\% outliers, $\mu=0, \beta=0, \sigma_w=4$ and $\sigma_\alpha=1$. Provider sample sizes are drawn from a discrete uniform distribution except for provider 1 with  (a) $n_1=25$, (b) $n_1=50$, (c) $n_1=100$ and (d) $n_1=125$. Outlier effects have equal probability of taking values of $\pm 4$. The x-axis represents the fixed value of $\alpha_1$. 	Dashed black lines represent the estimated probability when the true parameters $\sigma_w^2$ and $\sigma_{\alpha}^2$ are known. }
	\label{fig4_lin_out}
\end{figure}

To illustrate the robustness of EN compared to FERE, we also simulate $N=3,000$ providers with 5\% outliers. Still, we assume model (\ref{eq:linear_model}) for the majority of providers, with $\mu=0$, $\beta=0$, $\sigma_w=4$, and $\sigma_{\alpha}=1$ except that $\alpha_1$ is fixed at a value varying from 0 to $3.5\sigma_{\alpha}$. Half of the outliers have provider effect $\alpha_i = 4\sigma_{\alpha}$ and the other half $\alpha_i = -4\sigma_{\alpha}$, well outside the center of the true distribution for the majority of provider effects. Sample size $n_i$ is simulated from a uniform distribution on integers $\{10, 11, \cdots, 150\}$ for all providers except provider 1. For provider 1, its sample size $n_1 = 25, 50, 100, 125$. The empirical null distribution is smoothed using the methods in Section 3.2. The simulation is repeated 1000 times. Fig. \ref{fig4_lin_out} plots the proportion of times that provider 1 is flagged in the presence of 5\% outliers. In this setting, the EN method results in almost identical performance compared to the case when the true variance parameters $\sigma_{\alpha}$ and $\sigma_w$ are known (black dashed line), and is more robust than FERE, which  has lower flagging proportions due to over-estimation of the between-provider variance, especially when  $n_1$ is large.

\subsection{Simulation of the SMR}

We consider a realistic situation where providers are of different sizes and simulate survival outcomes that mimic the SMR example in Section \ref{ext_nonlinear}. A similar study where providers are of the same size is presented in the Supplementary Materials Section S1.

We generate $N=2000$ providers whose sample sizes are simulated from a uniform distribution on integers in $[10,200]$, and then fixed throughout. For the $j$th subject in the $i$th provider, the survival time $T_{ij}$ follows an exponential distribution with hazard $\lambda_{ij} = 0.1\times \exp\{ \alpha_i + X_{ij1}\beta_1 + X_{ij2}\beta_2 \}$,
where $\beta_1=1$ and $\beta_2=-1$, $\alpha_i \overset{iid}{\sim} N(0, 0.2^2)$ are the provider effects, and the covariates, $X_{ij1}$ and $X_{ij2}$, are independent $N(0,1)$ variables. The censoring time $C_{ij} \overset{iid}{\sim} Unif(10,30)$, which generates approximately 27\% censoring. We fit a Cox proportional hazards model with facilities as strata  and obtain  regression coefficient estimates for the covariates. The raw p-values and corresponding Z-scores are computed as described in Section \ref{ext_nonlinear}. We implemented the smoothed empirical null approach from Section \ref{en} with different numbers of groups $G = 5,20,40,60$. Linear regression models are fitted to the group-wise variance estimates of Z-scores, and weighted smoothing splines to the group-wise mean estimates. We repeat the simulation 500 times.

\begin{figure}
	\centering
	\includegraphics[width=0.7\textwidth]{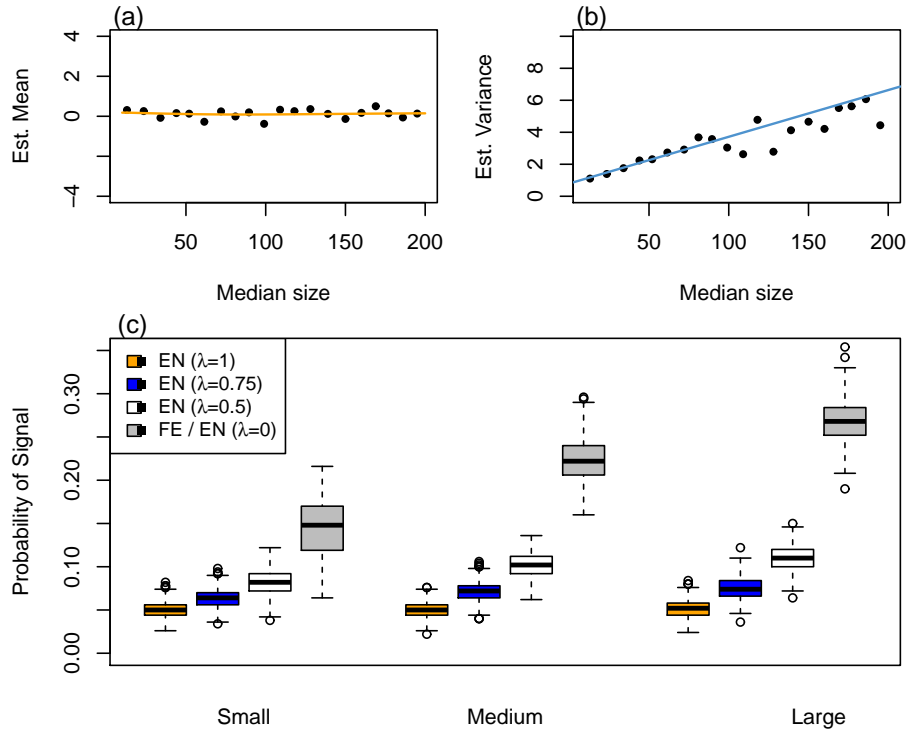}
	\caption{(a) Estimated mean function and (b) estimated variance function, by the smoothed empirical null, in one replication of the simulation with survival outcomes, with the number of groups $G = 20$. Black dots represent group-wise robust mean and variance estimates of  Z-scores. (c) Boxplots of empirical probability of signal summarized over 500 replications, all providers stratified into three groups by provider size. A proportion $\lambda = 0, 0.5, 0.75, 1$ of the between-provider variance is assumed to be due to incomplete risk adjustment, and the rest due to the quality of care. $\lambda = 0$ corresponds to the fixed-effects analysis.}
	\label{fig5_smoothEN}
\end{figure}

Fig.  \ref{fig5_smoothEN}(a)-(b) show the estimated mean and variance functions of Z-scores in one replication when the number of groups $G=20$. Results with different $G$  show no sensitivity to the selection of $G$ in the range 5 to 60 (Section S2 in the Supplementary Materials). Our proposed method captures the main features of the mean and variance functions while being smooth enough to provide consistent flagging rules within the considered range of provider size.

If one assumes that a proportion $\lambda$ of the between-provider variance is due to incomplete risk adjustment, then the methods of Section 3.4 can be used. Following the same simulation setup with survival outcomes above, we consider $\lambda = 0, 0.5, 0.75, 1$ and modify the allowed variance $\hat{\sigma}_{\lambda, i}^2$ in the reference distribution accordingly. Fig. \ref{fig5_smoothEN}(c) shows box plots of the proportion of times that providers are flagged in 500 replications, stratified into three groups. The FE approach, corresponding to $\lambda=0$, flags a provider if its one-sided p-value is less than 0.05, which results in over 25\% of the large providers and about 15\% of the small providers being signaled. In contrast, the EN approach that allows all variation ($\lambda=1$) has very stable flagging rates around 5\% for all three groups. The EN approach with the relaxed factor $\lambda$ can be viewed as a hybrid of the FE analysis and the EN allowing total variance, hence its flagging rates lie between the latter two. When $\lambda=0.5, 0.75$, the flagging rates increase somewhat with provider size, a feature inherited from FE.

\subsection{Mortality in U.S. dialysis facilities}

The Standardized Mortality Ratio (SMR) is used as a measure of mortality to profile dialysis facilities at CMS. More details on SMR can be found in Sections \ref{sec1} and \ref{ext_nonlinear}. Data were collected from 2012 to 2015, involving over half a million dialysis patients. In this analysis, we include 6,363 dialysis facilities with expected number of deaths of 3.0 or more.  The number of observed deaths ranges from 0 to 581, the number of expected deaths from 3.0 to 308.6, and the facility sizes ranges from 6.9 to 1569.8 patient-years. Facility size tertiles, defined by cut points at 156.9 and 302.8 patient-years, create three groups of small, medium and large facilities. 

Fig. \ref{fig2_smr} shows the histograms of FE Z-scores by stratum, the distribution of facility size and different flagging threshold lines for the stratified and the smoothed EN methods, and has been discussed in Section \ref{sec3}. Switching from the stratified EN to the smoothed version changes the flagging labels for some facilities. Nine facilities are flagged as ``worse than expected" by the stratified EN but not the smoothed EN, and 18 facilities  are flagged by the smoothed EN but not the other. A total of 367 facilities are flagged by both EN methods, and 5,968 facilities by neither. The FE approach results in 768 facilities being flagged (12.1\% of the total number), with 231 (10.9\%), 241 (11.4\%) and 296 (14.0\%) in the small, medium and large groups, respectively (percentages are with respect to the number of facilities in each group). The FE approach flags more large facilities than small ones, and the flagging rates exceed the target 5\% by large margins due at least in part to overdispersion. Using the smoothed EN, the flagging rates in the three groups are brought down to a more equitable level, 141 (6.6\%), 109 (5.1\%) and 123 (5.8\%). Due to possible existence of outliers, the empirical null based flagging rates are slightly larger than the target level.


\section{Conclusions and Discussion}

Besides the comments in Section 3.4, one may also choose a $\lambda$ value that would result in a certain proportion of providers being flagged, especially when there are constraints on resources made available for the review process or quality improvement program. This could also be accomplished by changing the nominal flagging rate $\rho$.

Generalization of RE and FERE methods to non-linear models, such as the logistic model or the Cox proportional hazards model, is complicated. For example, methods developed for assessing hospital readmission rates were based on an RE analysis of a hierarchical logistic model \citep{horwitz2011hospital}, and entailed complicated bootstrap techniques to assess significance. The EN method, however, generalizes immediately as described in Section 3.3, and is applied to a more complicated logistic model for hospital readmissions of dialysis patients in \citep{he2013evaluating}.

In the EN approach, we have assumed that there are a large number of providers to be profiled and that the central part of the histogram of Z-scores is well described by a normal distribution. These two assumptions are satisfied in many applications. When the number of facilities is much smaller, it is important to take into account uncertainty in the estimation of the empirical null distribution. Also, in some instances, it may be useful to use a transformation of the basic measure in order to achieve approximate normality of the empirical null. In other cases, there may be situations where a non normal distribution that incorporates skewness, for example, is more appropriate. These are areas where additional work is needed.

As noted earlier, an FE analysis to estimate $\beta$ and then use of an offset is one approach to correct for the confounding between covariates and provider effects in the RE method. An alternative approach is to include both a within and a between regression coefficient in the model
\[   Y_{ij}=\mu+\alpha_i+\beta_w^T(X_{ij}-\bar X_i)+ \beta_b^T\bar X_i + \epsilon_{ij}    \]
as described in \citet{neuhaus1998between}. A feature which sometimes arises is that the between-provider regression coefficient, $\beta_b \neq \beta_w$, in which case the within coefficient does not make a full adjustment for the covariates under consideration. This would arise, for example, if even having adjusted for a variable like race, one found that there was still a difference between providers according to the racial mix that they treat. Such situations require careful consideration, and a discussion of issues associated with this can be found in \citet{kalbfleisch2018does}.

Bayesian methods are computationally attractive and have often been used in profiling. These impose distributional assumptions on the random provider effects, often with hyperparameters in full Bayesian methods \citep{normand1997statistical, racz2010bayesian, normand2007statistical}. However, there still remains confusion as to nonunified criteria for identifying outlying providers based on posterior distributions. For instance, \citet{racz2010bayesian} assessed whether a pre-determined norm lies between the posterior percentiles, e.g. 2.5\% and 97.5\%, of a provider effect in a hierarchical logistic regression model. \citet{normand1997statistical} considered the posterior probability of the excess expected mortality (the difference between the expected mortality under a provider's own regression coefficients and that averaged over the provider-specific parameters) being larger than a benchmark, and the posterior probability of the adjusted mortality of a reference patient greater than that for similar patients in all providers in the same sample. Often, extremeness of the observed mortality have also been assessed based on the posterior predictive distrbution through replications  \citep{normand2007statistical}. Based on a hierarchical logistic regression model, Bayesian intervals can be constructed for standardized readmission measures via bootstrapping from a normal approximation to the posterior distribution of provider effects in an approach similar to the RE methods discussed in this paper.   As with RE methods, it is worthwhile to take measures with a Bayesian approach to ensure fair assessments of providers of all sizes.

A causal inference framework provides a promising but challenging approach to profiling health care providers. In general, the existence of unmeasured confounders poses difficulties in the inference on providers' performance. \citet{spertus2016assessing} implemented augmented inverse probability weighting \citep{robins1994estimation} and targeted maximum likelihood estimation \citep{van2006targeted}  under a causal inference framework for profiling, coupled with elastic net for variable selection.  \citet{spertus2016assessing}  discussed using instrumental variables if one has strong reasons to assume the underlying causal mechanism. As is often the case, for causal inference, cautions are needed in connection with many commonly made but non-verifiable assumptions.

\section*{Software}

The R code for implementing the empirical null with simulation examples in this paper has been made available at \url{https://github.com/luxia-bios/Empirical-Null}.

\section*{Supplementary Material}

Supplementary material will be available online at
\url{http://biostatistics.oxfordjournals.org}.

\section*{Acknowledgements}

This work was supported in part by The Centers for Medicare and Medicaid Services  [contract number HHSM-500-2008-000211], although the opinions expressed in this article do not necessarily reflect those of the CMS or the U.S. government.  \textit{Conflict of Interest:} None.

\bibliographystyle{biorefs}
\bibliography{refs.bib}


\newpage


\end{document}